\documentclass[twocolumn,trackchanges,dvipdfmx]{aastex631}
\usepackage[dvipdfmx]{graphicx}
\usepackage{comment}

\def\kms{\hbox{km s$^{-1}$}}
\def\VLSR{\hbox{$V_{\rm LSR}$}}

\def\sun{\hbox{$_{\odot}$}}

\def\Hone{\hbox{H\,{\scriptsize I}}}

\received{}
\revised{}
\accepted{}

\shorttitle{Millimeter-wave CO and SiO Observations toward the BVF CO 16.134--0.553: a Smith cloud scenario?}
\shortauthors{Yokozuka et al.}

\begin{document}

\title{Millimeter-wave CO and SiO Observations toward the Broad-velocity-width Molecular Feature CO 16.134--0.553: a Smith cloud scenario?} 

\correspondingauthor{Hiroki Yokozuka}
\email{gackt4869akai@keio.jp}

\author[0000-0002-0786-7307]{Hiroki Yokozuka}
\affil{School of Fundamental Science and Technology, Graduate School of Science and Technology, Keio University, 3-14-1 Hiyoshi, Kohoku-ku, Yokohama, Kanagawa 223-8522, Japan}

\author[0000-0002-5566-0634]{Tomoharu Oka}
\affiliation{School of Fundamental Science and Technology, Graduate School of Science and Technology, Keio University, 3-14-1 Hiyoshi, Kohoku-ku, Yokohama, Kanagawa 223-8522, Japan}
\affiliation{Department of Physics, Institute of Science and Technology, Keio University, 3-14-1 Hiyoshi, Kohoku-ku, Yokohama, Kanagawa 223-8522, Japan}
\author[0000-0002-1663-9103]{Shiho Tsujimoto}
\affiliation{Department of Physics, Institute of Science and Technology, Keio University, 3-14-1 Hiyoshi, Kohoku-ku, Yokohama, Kanagawa 223-8522, Japan}
\author{Yuto Watanabe}
\affiliation{School of Fundamental Science and Technology, Graduate School of Science and Technology, Keio University, 3-14-1 Hiyoshi, Kohoku-ku, Yokohama, Kanagawa 223-8522, Japan}
\author[0000-0002-0786-7307]{Miyuki Kaneko}
\affiliation{School of Fundamental Science and Technology, Graduate School of Science and Technology, Keio University, 3-14-1 Hiyoshi, Kohoku-ku, Yokohama, Kanagawa 223-8522, Japan}

\begin{abstract}
We report the results of the CO {\it J}=1--0 and SiO {\it J}=2--1 mapping observations towards the broad-velocity-width molecular feature CO 16.134--0.553 with the Nobeyama Radio Observatory 45 m telescope.  The high quality CO map shows that the 5-pc size broad-velocity-width feature bridges two separate velocity components at $\textit V_{\rm{LSR}}\! \simeq\! 40$ \kms\ and $65$ \kms\ in the position-velocity space.  The kinetic power of CO 16.134--0.553 amounts to $7.8\!\times\! 10^2$ $L_{\sun}$ whereas no apparent driving sources were identified.  Prominent SiO emission was detected from the broad-velocity-width feature and its root in the $\textit V_{\rm{LSR}}\! \simeq\! 40$ \kms\ component.  In the CO Galactic plane survey data, CO 16.134--0.553 appears to correspond to the Galactic eastern rim of a 15-pc diameter expanding CO shell.  An $1\arcdeg$-diameter \Hone\ emission void and $4\arcdeg$-long vertical \Hone\ filament were also found above and below the CO shell, respectively.  We propose that the high-velocity plunge of a dark matter subhalo with a clump of baryonic matter was responsible for the formation of the \Hone\ void, CO 16.134--0.553/CO shell, and the \Hone\ filament.  
\end{abstract}
\keywords{Galaxy: disk --- ISM: clouds --- ISM: molecules}

\section{Introduction} 
\label{sec:intro}
CO 16.134--0.553 is a broad-velocity-width molecular feature (BVF) found in the Galactic plane with a spatial size of $3\!\times\! 4$ pc$^2$ and a velocity width of $\sim\! 40$ \kms\ \citep{Yokozuka21}.  This was discovered through a systematic, unbiased search for compact ($d\! <\! 10$ pc) broad-velocity-width ($\Delta V\! >\! 5$ \kms ) features in the CO {\it J}=1--0 Galactic plane survey data obtained with the Nobeyama Radio Observatory (NRO) 45 m telescope \citep[the FUGIN survey;][]{Umemoto17}.  The virial theorem mass of CO 16.134--0.553, $M_{\rm VT}\!\simeq\! 2\!\times\! 10^5\, M_{\sun}$, is $1.5$ orders of magnitude higher than the gas mass, $M\!\simeq\! 7\!\times\! 10^3\, M_{\sun}$.  It exhibits huge kinetic energy, $E_{\rm kin}\!\sim\! 10^{49}$ erg, and a short dynamical time, $t_{\rm dyn}\!\simeq\! 1\!\times\! 10^5$ yr, which results in a kinetic power as high as $\sim\! 10^3\, L_{\sun}$.  
 
This is only one out of 58 BVFs that has no counterpart either in the radio continuum (\citealp[1.4 GHz:][]{Beuther16, Wang20};  \citealp[10 GHz:][]{Handa87}), far-infrared \citep[60--110 $\mu$m:][]{Doi15, Takita15}, or mid-infrared \citep[3.4 and 4.6 $\mu$m:][]{Wright10} images \citep{Yokozuka21}.  Referring to the infrared luminosity ($L_{\rm IR}$) vs. kinetic power ($P_{\rm kin}$) plot, the 57 BVFs other than CO 16.134--0.553, were likely driven by protostellar outflows \citep{Beuther02, Maud15, Li18, Zhang20}.  Despite the huge kinetic power of CO 16.134--0.553, its driving source is currently unidentified.  This situation is quite similar to that of many high-velocity dispersion compact clouds (HVCCs) found in the central molecular zone of our Galaxy \citep[e.g.,][]{Oka22}

To unveil the nature and origin of CO 16.134--0.553, we conducted follow-up observations using the Nobeyama Radio Observatory (NRO) 45 m telescope (\S 2).  This paper presents the newly obtained high-quality CO {\it J}=1--0 line images and reports on the detection of SiO {\it J}=2--1 line emission from this peculiar BVF (\S 3).  We also revisited the FUGIN CO and \Hone\ survey data sets to determine the relationship to the large-scale structures (\S 4).  Based on these data sets, we proposed a formation scenario for CO 16.134--0.553 (\S 5).  The study is summarized in the last section (\S 6).  The distance to CO 16.134--0.553 was assumed to be $D\!=\! 4.0$ kpc \citep{Yokozuka21}.

\section{Observations} 
We observed CO {\it J}=1--0 (115.271 GHz) and SiO {\it J}=2--1 (86.847 GHz) lines using the NRO 45 m telescope.  A $382\arcsec\!\times\! 382\arcsec$ area, which covers the entire CO 16.134--0.553, was mapped in 17--18 February 2020.  The four-beam receiver system on the 45-m telescope \citep[FOREST;][]{Minamidani16} with the spectral analysis machine on the 45-m telescope \citep[SAM45;][]{Kuno11, Kamazaki12} system were used in the observations.  The half-power beamwidths (HPBWs) of the telescope with FOREST were approximately $14\arcsec$ and $19\arcsec$ at 115 GHz and 86 GHz, respectively.  We used the SAM45 spectrometer in the 1 GHz bandwidth ($244.14$ kHz resolution) mode.  This frequency resolution corresponds to $1$ \kms\ velocity resolution at $115$ GHz.  The system noise temperature ($T_{\rm sys}$) ranged from 100 to 900 K during the CO observations, and from 100 to 150 K during the SiO observations.  The telescope pointing accuracy was checked and corrected every 2 hr by observing the SiO maser source VX-Sgr at 43 GHz.  The pointing accuracy was maintained within $3\arcsec$ in both azimuth and elevation.  The intensity calibration of the antenna temperature was accomplished using the standard chopper-wheel method.

The obtained data sets were reduced using the {\it NOSTAR} reduction package.  Linear fittings were used to subtract the baseline offsets from all the obtained spectra.  The maps were convolved using the Bessel--Gaussian function and resampled onto a $7\farcs 5\!\times\! 7\farcs 5\!\times\! 2\, \kms$ regular grid.  We scaled the antenna temperature (${T_{\rm A}^*}$) by multiplying it with $1/\eta_{\rm MB}$ to obtain the main-beam temperature ($T_{\rm MB}$). 

\begin{figure}[htbp]
\includegraphics[width=80 mm]{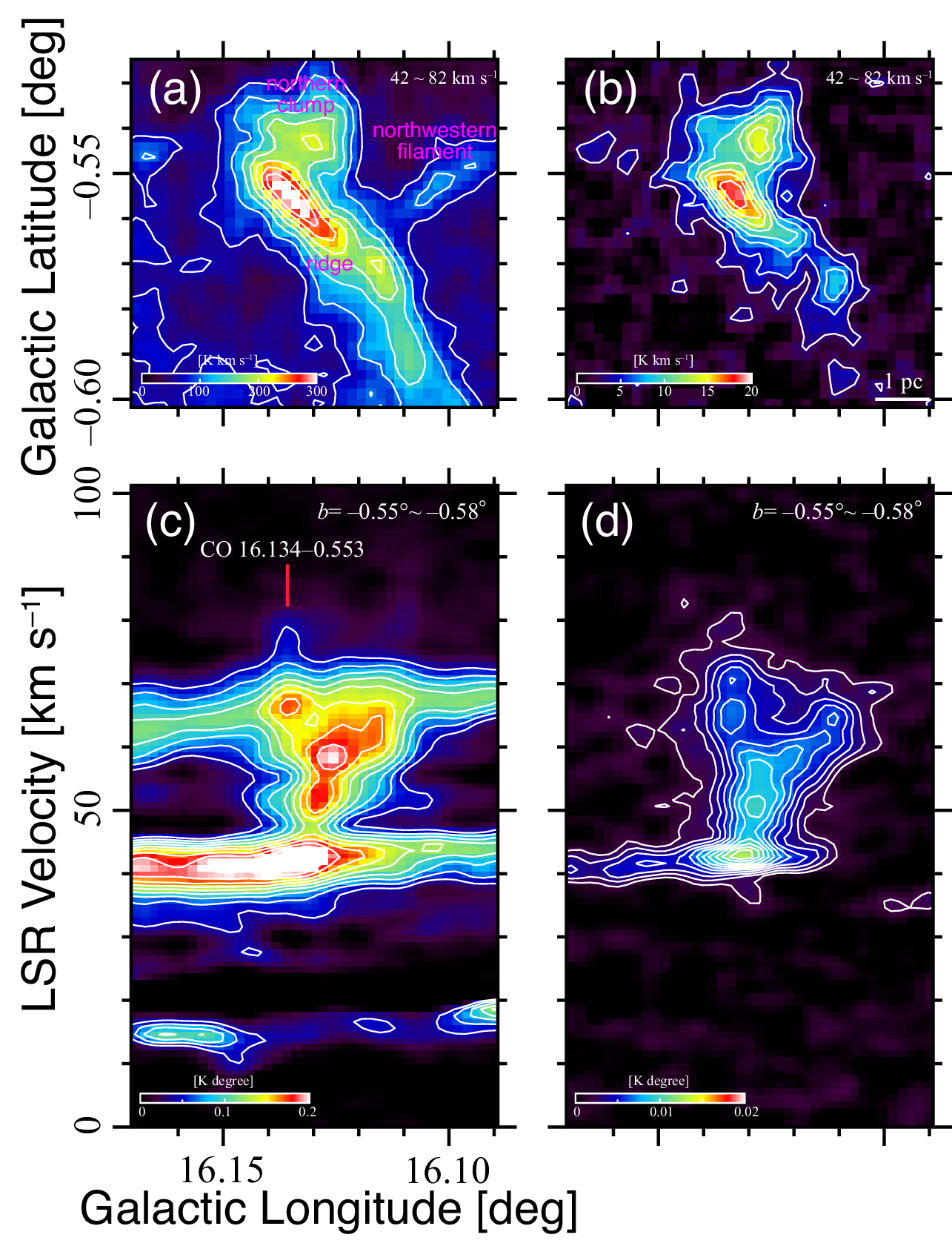}
\centering
\caption{(a)(b) Maps of the velocity-integrated CO {\it J}=1--0 and SiO {\it J}=2--1 line emission, respectively.  The velocity range for the integration is $\VLSR\! =\! 42\mbox{--}82$ \kms .  Contours are drawn at $35$ K \kms\ and $2$ K \kms\ intervals for CO and SiO, respectively.  (c)(d) Longitude-velocity maps of the CO {\it J}=1--0 and SiO {\it J}=2--1 line emission, respectively.  Each emission was integrated latitudes from $b\! =\!-0\fdg 55$ to $-0\fdg 58$.  Contours are drawn at $0.02$ K\arcdeg\ and $0.001$ K\arcdeg\ intervals for CO and SiO, respectively. }
\label{fig1}
\end{figure}

\begin{figure*}[htbp]
\includegraphics[width=145 mm]{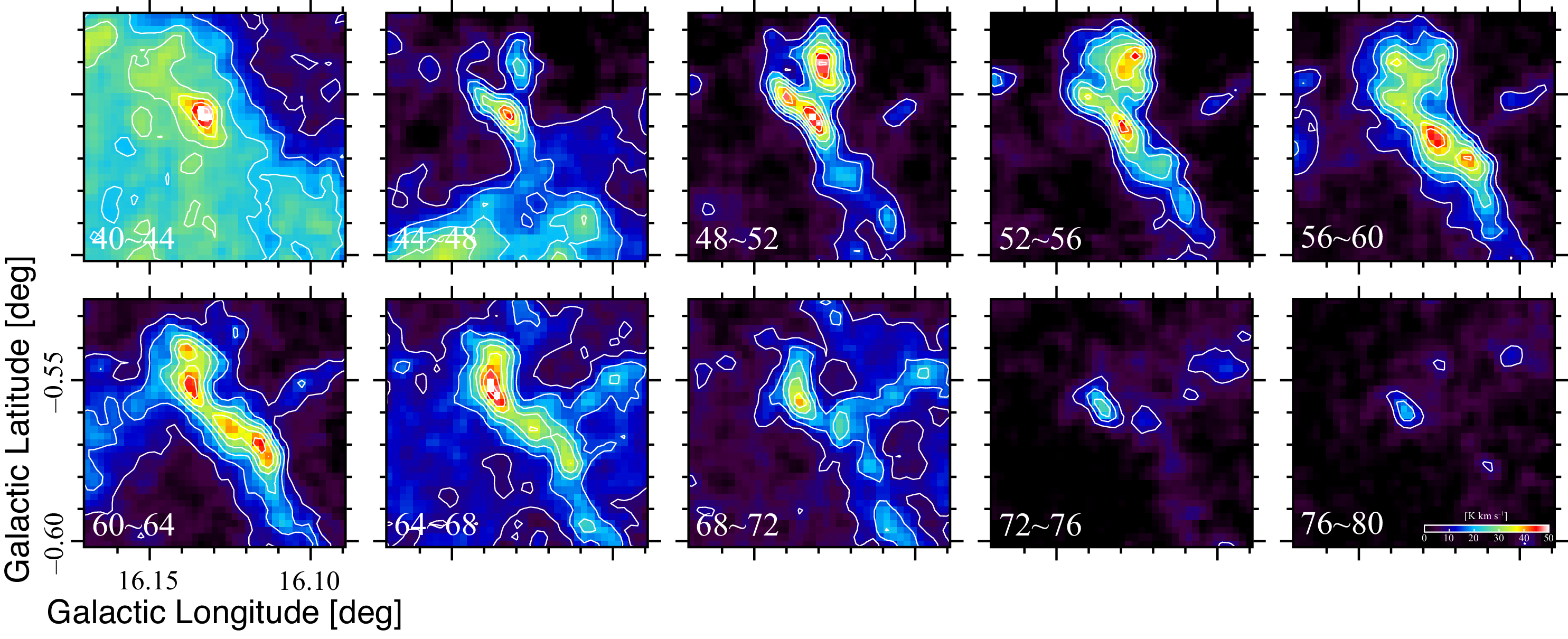}
\centering
\caption{CO {\it J}=1--0 integrated intensity channel maps calculated over a 4 \kms\ LSR velocity width.  The intensity unit is K \kms .  A pair of numbers in the bottom-left of each panel denotes the velocity range of integration.  Contours are drawn at $6.5$ K \kms\ intervals.}
\label{fig2}
\end{figure*}

\begin{figure*}[htbp]
\includegraphics[width=145 mm]{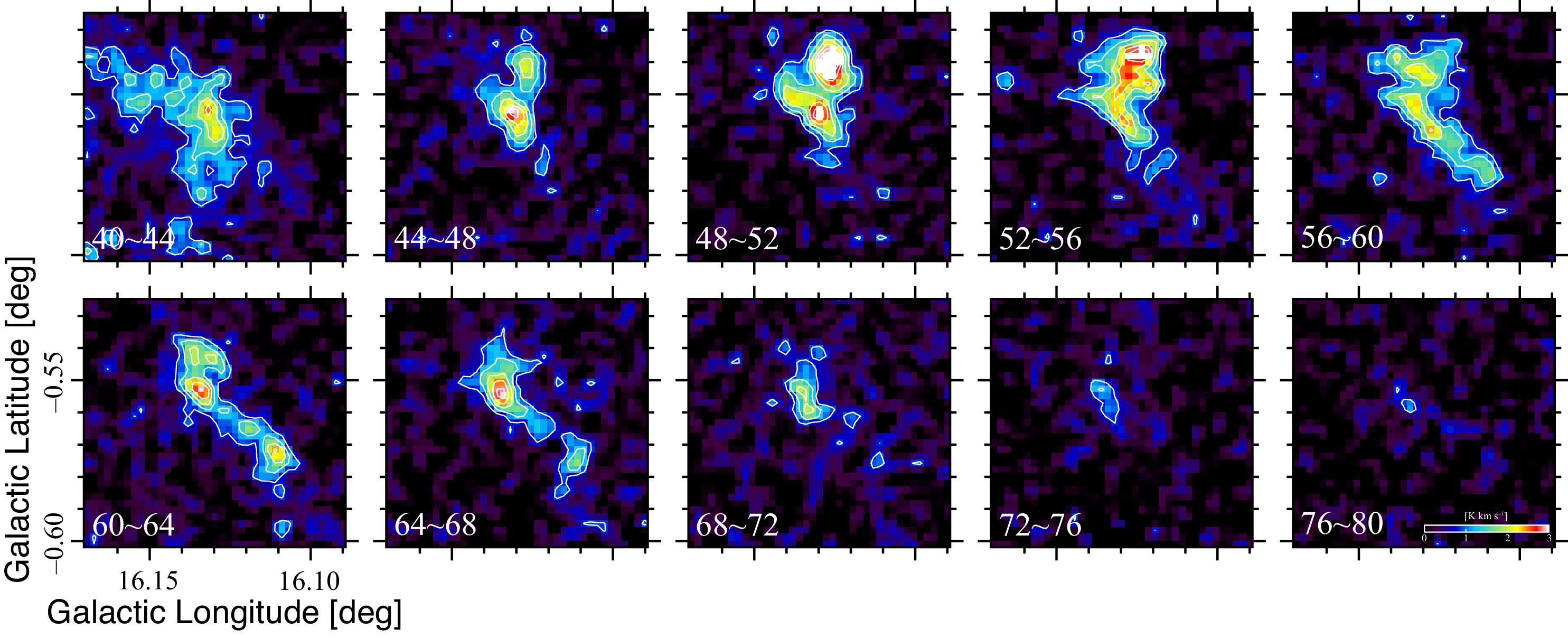}
\centering
\caption{SiO {\it J}=2--1 integrated intensity channel maps calculated over a 4 \kms\ LSR velocity width.  The format is the same as Figure 2.  Contours are drawn at $0.5$ K \kms\ intervals.}
\label{fig3}
\end{figure*}

\section{Results} 
\subsection{Morphology and Kinematics}
Figure 1ac shows the velocity-integrated and longitude-velocity ({\it l--V}) maps of CO {\it J}=1--0 line emission.  CO 16.134--0.553 comprises of a 5-pc length ridge elongated from Galactic southwest to northeast (hereafter referred to as ``ridge") and a 2-pc size clump in the north of the ridge (``northern clump").  The ridge turns southward at $(l, b)\!\simeq\! (16\fdg 11, -0\fdg 58)$.  A faint, filamentary structure (``northwestern filament") elongates toward northwest from the center  of the ridge.  In the {\it l--V} plane, the ridge spreads in a fan-shape toward higher velocities, and connects two spatially extended emission components at $\VLSR\!\simeq\! 40$ \kms\ and $65$ \kms .  The $40$ \kms\ component corresponds to molecular gas in the Norma arm, while the $65$ \kms\ component is not associated with the known Galactic spiral arm.  Faint high-velocity features appear to penetrate the $65$ \kms\ component reaching $\VLSR\!\simeq\! 80$ \kms . 

Prominent SiO {\it J}=2--1 line emission was detected from the ridge and northern clump of CO 16.134--0.553 (Figure 1bd).  The SiO emission is less spatially extended than the CO emission.  The southern half of the ridge is faint and clumpy.  In the {\it l--V} plane, the ridge bifurcates at $\VLSR\!\simeq\! 55$ \kms , forming a V-shape which traces the edge of the fan-shape in the CO {\it l--V} map.  The spatially extended $40$ \kms\ component is barely visible in the SiO {\it l--V} map, while the $65$ \kms\ component appears to be absent.  The positive high-velocity end of SiO emission also reaches to $\VLSR\!\simeq\! 80$ \kms .

\subsection{Velocity Channel Maps}
Figures 2 and 3 show the velocity-channel maps of the CO {\it J}=1--0 and SiO {\it J}=2--1 line emissions.  In the CO velocity channel maps, the ridge of CO 16.134--0.553 comprises several clumps of 1--2 pc in sizes .  The northern clump appears in the velocity range of $\VLSR\!=\! 44\mbox{--}64$ \kms , and merges with the ridge at the higher velocities.  The highest CO intensity was observed in the northern clump at $(l, b, \VLSR )\!=\! (16\fdg 13, -0\fdg 54, 51\,\kms)$.  The $40$ \kms\ component and faint $65$ \kms\ component spread over the frames at $\VLSR\!=\! 40\mbox{--}44$ \kms\ and $\VLSR\!=\! 64\mbox{--}68$ \kms\ channels, respectively.  The northwestern filament appears at $\VLSR\!=\! 64\mbox{--}72$ \kms .  A pair of faint, compact high-velocity features originate from the northwestern filament at $(l, b)\!\sim\! (16\fdg 10, -0\fdg 55)$ on both velocity sides.  The velocity extent of this pair of compact high-velocity features is $\VLSR\!\simeq\! 48\mbox{--}80$ \kms .  

The SiO velocity channel maps (Figure 3) delineate the less-extended spatial distribution of SiO emission.  The northern clump and the northern two-thirds of the ridge are prominent in the SiO maps.  The highest SiO intensity was observed at the center of the northern clump, with almost the same {\it l--b--V} position as that of the CO emission peak , $(l, b, \VLSR )\!=\! (16\fdg 13, -0\fdg 54, 50\,\kms)$.  The $40$ \kms\ component in the SiO map is less spatially extended and distributed around the northern part of the ridge.  The $65$ \kms\ component, the northwestern filament, and the southern one-thirds of the ridge are absent from the SiO maps.

\section{Discussion} 
\subsection{Physical Parameters}
Using the newly obtained CO {\it J}=1--0 data in the range of $16\fdg 10\!\leq\! l\!\leq\! 16\fdg 15$, $-0\fdg 60\! \leq\! b\!\leq\! -0\fdg 52$, $42\,\kms\,\! \leq\VLSR\!\leq\! 82\,\kms$, we recalculated the physical parameters of CO 16.134--0.553.  The size parameter and velocity dispersion \citep[definitions in][]{Solomon87} were calculated as $S\! =\! 1.2$ pc and $\sigma_{V}\! =\! 8.5$ \kms , respectively.  These yield a dynamical time as $t_{\rm dyn}\!=\! 1.4\!\times\! 10^{5}$ yr.  Assuming the local thermodynamic equilibrium (LTE), the molecular gas mass ($M_{\rm LTE}$) of CO 16.134--0.553 was obtained as $5.1\!\times\! 10^3 M_{\sun}$.  Here, we employed an optical depth of $\tau\! =\! 3.6$ and excitation temperature of $T_{\rm{ex}}\! =\! 18$ K, which were derived from the FUGIN CO and $^{13}$CO {\it J}=1--0 data using the equations (1)--(2) in \citet[][]{Oka98} with [CO]/[$^{13}$CO]$\! =\! 60$.  The kinetic energy was calculated using the definition, $E_{\rm kin}\! =\!(3/2)M_{\rm LTE}\sigma_{V}^{2}$, and obtained as $1.3\!\times\! 10^{49}$ erg.  The kinetic energy and the dynamical time calculated above correspond to the kinetic power ($P_{\rm kin}\!\equiv\! E_{\rm kin}/t_{\rm dyn}$) as high as $7.8\!\times\! 10^{2}$ $L_{\sun}$ ($=\! 3.0\!\times\! 10^{36}$ erg).  This is at least one order of magnitude higher than that of bipolar outflows from massive protostars observed to date \citep[e.g.,][]{Maud15}.  Moreover, the absence of infrared counterpart resulted in a large departure of CO 16.134--0.553 from the infrared luminosity to kinetic power ($L_{\rm IR}$-$P_{\rm kin}$) trend of protostellar outflows \citep{Yokozuka21}.

\begin{deluxetable}{lc}\label{tab:ratio}
\tablecaption{SiO {\it J}=2--1/CO  {\it J}=1--0 intensity ratios }
\tablewidth{0pt}
\tablehead{
\colhead{Component} & \colhead{Ratio$^{a}$} 
}
\startdata
Ridge &  $0.038\pm 0.005$  \\ 
Northern clump & $0.059\pm 0.008$ \\ 
Northwestern filament & $0.017\pm 0.002$ \\ 
40 \kms\ component & $0.0034\pm 0.0004^{b}$ \\ 
65 \kms\ component & $0.0066\pm 0.0010^{b}$ \\ 
\hline
IRAS 00338+6312 & $0.028^{c}$ \\ 
IRAS 00494+5617 & $0.018^{c}$ \\ 
CMZ & $0.009^{d}$ \\ 
\enddata
\tablecomments{(a) Uncertainties come from the absolute calibration error of the data which were estimated from the intensity reproducibility during observations ($\simeq\! 10$\%).   
(b) Calculated by using pixels with $T_{\rm MB}({\rm SiO})\!\leq\!0.15$ K.  Velocity ranges for integration were $\VLSR\!=\!38$--$48$ \kms\ and $60$--$75$ \kms\ for 40 \kms\ component and 65 \kms\ component, respectively.  
(c) Calculated from peak temperatures at each protostellar outflow presented in \citet{Harju98} and \citet{Snell90}.  
(d) Typical value in the central molecular zone of our Galaxy (S. Takekawa et al. 2024, in preparation) }
\end{deluxetable}

\subsection{Shock Signatures}
Gas-phase SiO is a well-established probe for strong shock in interstellar space \citep[e.g.,][]{Ziurys82}, and the critical density of SiO {\it J}=2--1 line is as high as $\sim\! 10^{5}$ cm$^{-3}$ \citep[e.g.,][]{Huang22}.  Thus, the detection of widespread, broad-velocity-width SiO {\it J}=2--1 emission from CO 16.134--0.553 indicates the presence of dense molecular gas which is affected by strong interstellar shock.  Here we referred to the SiO {\it J}=2--1/CO  {\it J}=1--0 intensity ratio  (hereafter $R_{\rm SiO/CO}$) to assess the abundance of dense/shocked gas (Table \ref{tab:ratio}).  Ratios were calculated from velocity-integrated line intensities in units of kelvin kilometers per second.  Apparently, $R_{\rm SiO/CO}$ is determined with high significance across the whole of CO 16.134--0.553, being comparable to or even higher than those found at protostellar outflows (e.g., IRAS 00338+6312, IRAS 00494+5617).  The ratio is the highest in the northern clump and is slightly lower in the northwestern filament.  The 40 \kms\ component exhibits a very low $R_{\rm SiO/CO}$, while the 65 \kms\ component exhibits a value similar to the typical value in the central molecular zone (CMZ) of our Galaxy (S. Takekawa et al. 2024, in preparation).  These results indicate that CO 16.134--0.553, and possibly the 65 \kms\ component, have recently been experienced the passage of a strong interstellar shock. 


\begin{figure*}[htbp]
\includegraphics[width=110 mm]{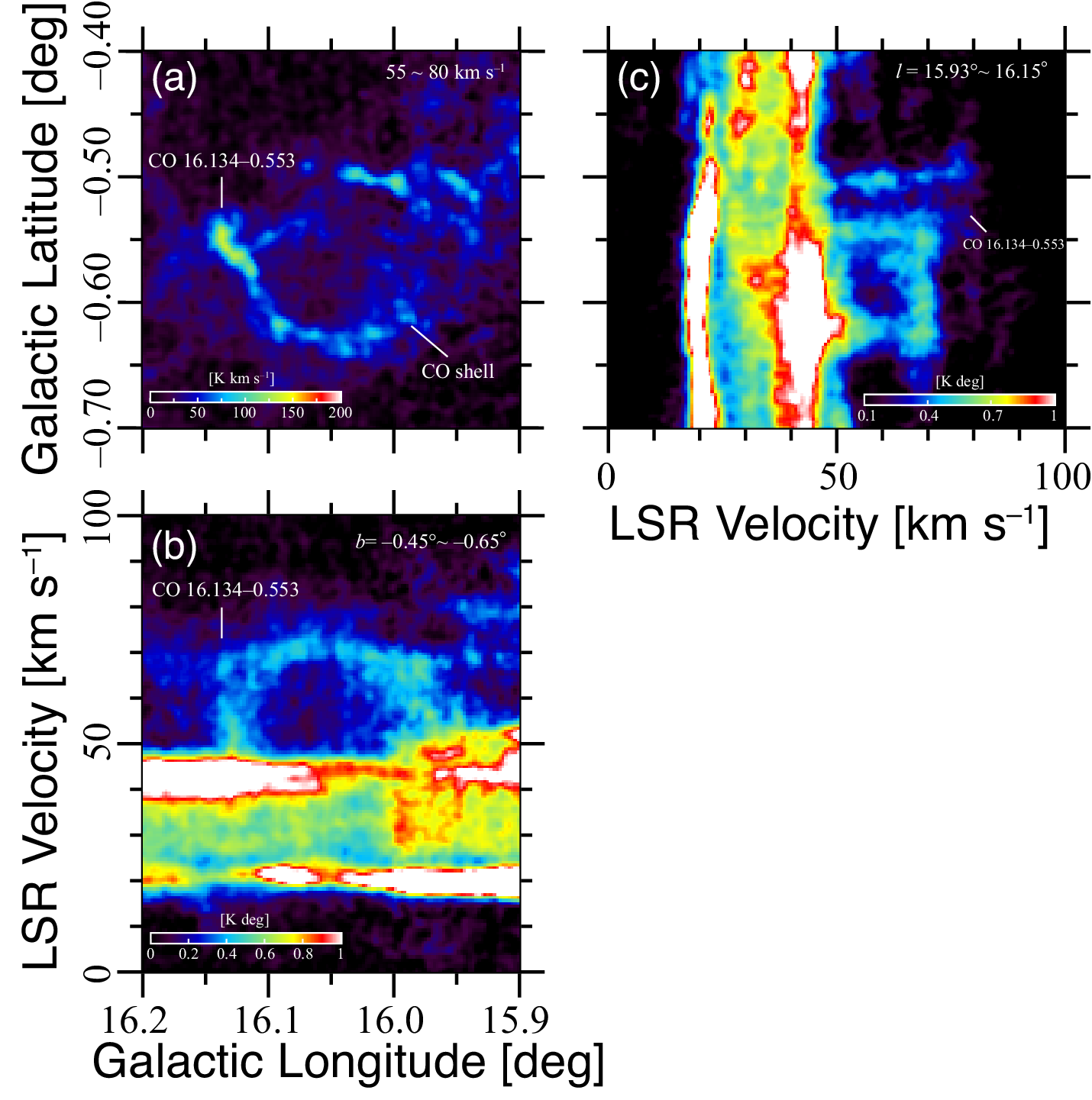}
\centering
\caption{(a) Map of velocity-integrated CO {\it J}=1--0 data obtained via the FUGIN survey.  The velocity range for the integration is from $\VLSR\! =\! 55$ to $80$ \kms .   (b) Map of latitude-integrated CO {\it J}=1--0 line emission.\ The latitude range for integration is from $b\! =\! -0\fdg 65$ to $-0\fdg 45$.  (c) Map of the longitude-integrated CO {\it J}=1--0 line emission.\ The longitude range for integration is from $l\! =\! 15\fdg 93$ to $16\fdg 15$.}
\label{fig4}
\end{figure*}

\begin{figure*}[htbp]
\includegraphics[width=110 mm]{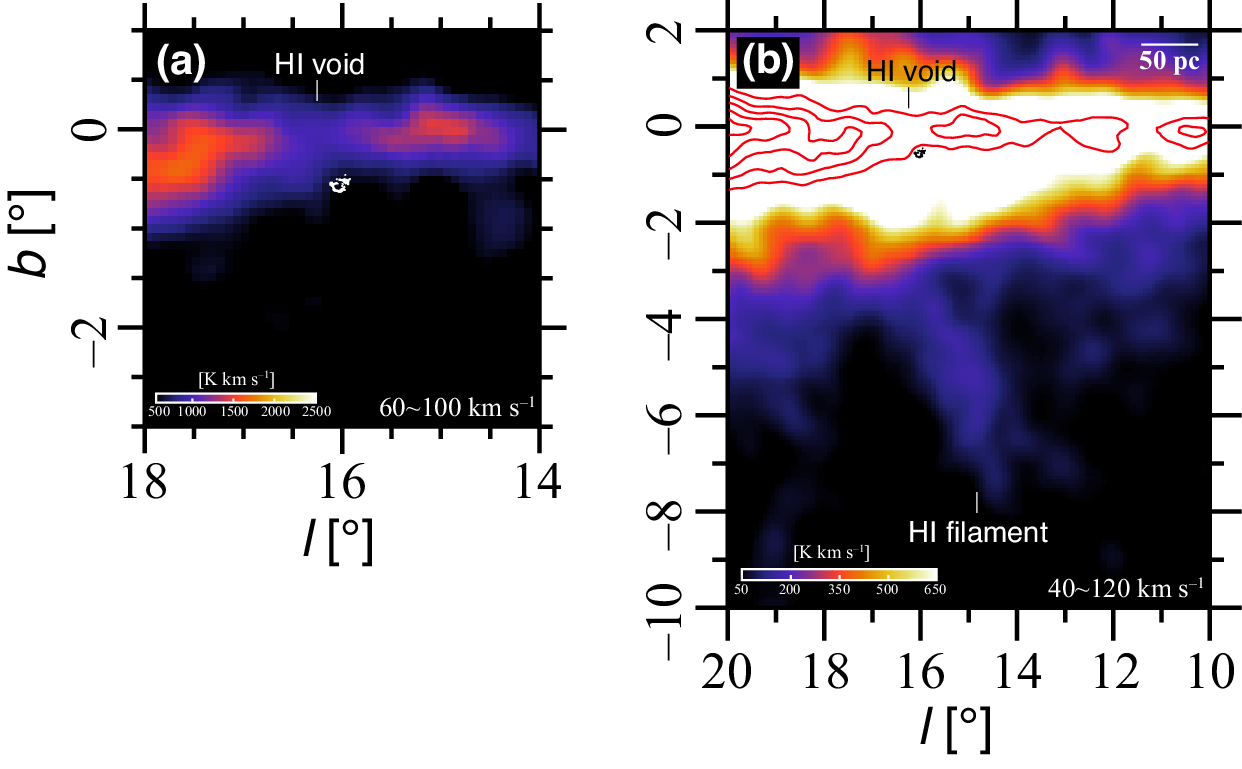}
\centering
\caption{Maps of \Hone\ 21 cm line emission around the CO shell denoted by white or black contours. The intensity unit is kelvin kilometers per second. (a) A $4\arcdeg\!\times\! 4\arcdeg$ map of \Hone\ emission integrated over velocities between $\VLSR\! =\! 60$ \kms\ and $100$ \kms .  (b) A $10\arcdeg\!\times\! 12\arcdeg$ map of \Hone\ emission integrated over velocities between $\VLSR\! =\! 40$ \kms\ and $120$ \kms .  Red contours show \Hone\ emission integrated from $60$ \kms\ to $100$ \kms\ at levels 100, 150, and 200 K \kms . }
\label{fig5}
\end{figure*}

\subsection{Large-scale Views}
The presence of the spatially-extended 65 \kms\ component and evidence for a widespread, strong shock indicate that a large-scale process is responsible for the acceleration of CO 16.134--0.553.  Again, we referred to CO {\it J}=1--0 data from the FUGIN survey \citep[][]{Umemoto17} to reexamine the large-scale distribution and kinematics of molecular gas around CO 16.134--0.553.  In the velocity-integrate CO map (Figure \ref{fig4}a), we noticed a shell-like structure with a diameter of $\sim\! 15$ pc.  Here, CO 16.134--0.553 appears to define the eastern edge of this 15-pc diameter CO shell (hereafter, ``CO shell").  The 65 \kms\ component exhibits a spatial extent as large as that of the CO shell, while the 40 \kms\ component spreads over the frame (Figure \ref{fig4}ab).  Several broad-velocity-width features, including CO 16.134--0.553, connect the 40 \kms\ and 65 \kms\ components (Figure \ref{fig4}bc).  The total molecular gas mass of the CO shell is $\sim\! 2\!\times\! 10^4\, M_{\sun}$ (assuming $X_{\rm CO}\!=\! 2\!\times\! 10^{20}\,{\rm cm}^{-2}(\mbox{K km s}^{-1})^{-1}$), and the kinetic energy amounts to $\sim\! 10^{51}$ erg.  The situation of CO 16.134--0.553/CO shell is very similar to that of the Smith cloud \citep[][]{Smith63}, which is a high-velocity cloud plunging into the \Hone\ disk of our Galaxy \citep{Lockman84, Lockman08}; however, the scale (size, mass, and kinetic energy) is far smaller (detailes in section \ref{sec:scenario}).  

We also referred to \Hone\ 21 cm line data obtained from the \Hone\ $4\pi$ survey \citep{HI4P}.  Figure \ref{fig5} shows two large-scale maps of the \Hone\ 21 cm line emission with different area sizes and velocity ranges for integration.  We noticed a local decrease of emission at velocities from $\VLSR\!=\! 60$ \kms\ to $100$ \kms\ in the intense \Hone\ disk of the Galaxy (Figure \ref{fig5}a).  This ``\Hone\ void" is located at $\sim\! 0\fdg 5$ above the CO shell, having a longitudinal size of $\sim\! 1\arcdeg$ ($=\! 70$ pc at a distance of $4$ kpc).  Far below the \Hone\ void/CO shell, we found a large filament with faint \Hone\ emission at velocities from $\VLSR\!=\! 40$ \kms\ to $120$ \kms\ (Figure \ref{fig5}b).  This ``\Hone\ filament" has the apparent length and width of $\sim\! 4\arcdeg$ and $\sim\! 1\arcdeg$, respectively, corresponding to $280$ pc and $70$ pc at $4$ kpc, respectively.  This extends from the Galactic \Hone\ disk, from the position of the \Hone\ void/CO shell, toward negative Galactic latitudes.  The total atomic gas mass of the \Hone\ filament is $\sim\! 10^5\, M_{\sun}$ (using the method in \citealt{Dickey90}).  This situation may indicate that the high-velocity plunging of a massive object from the Galactic halo was responsible for the formation of the \Hone\ void/CO shell (CO 16.134--0.553)/\Hone\ filament.

\begin{figure}[tbh]
\includegraphics[width=60 mm]{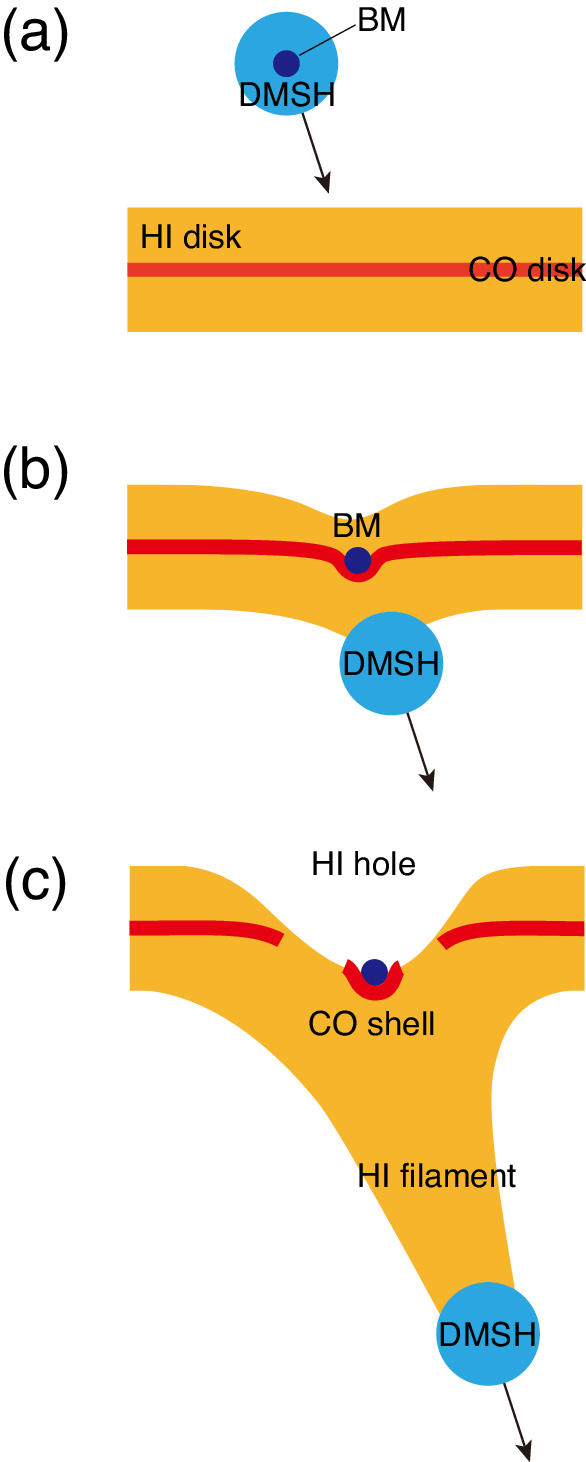}
\centering
\caption{Schematic image of the Massive Halo Object plunging scenario.  The cyan and blue filled circles show a dark matter subhalo (DMSH) and a clump of baryonic matter (BM), respectively.  The orange area represents the atomic gas (\Hone ) distribution, and the thick red line shows the molecular gas (CO) distribution.  }
\label{fig6}
\end{figure}

\subsection{Plunging Massive Halo Object Hypothesis}\label{sec:scenario}
Based on the observational facts presented above, we propose the following scenario (Figure \ref{fig6}):  
\begin{enumerate}
\setlength{\parskip}{0pt} 
\setlength{\parsep}{0pt} 
\setlength{\itemsep}{1pt} 
\item A small dark matter subhalo (DMSH) accompanied by a baryonic matter clump (BM) plunged into the Galactic disk. 
\item The BM was stopped by the CO disk forming the CO shell and CO 16.134--0.553, while the DMSH penetrated the Galactic disk. 
\item The DMSH formed the \Hone\ filament by the gravitational attraction, leaving the \Hone\ void in the Galactic disk.  
\end{enumerate}
This scenario was invoked by an analogy from the Smith cloud \citep{Lockman08} and theoretical works that simulate a DMSH with BM impacting the Galactic disk (e.g., \citealt{Bekki2006}; \citealt{tepper}; \citealt{Shah2019}).  The plunging scenario naturally explains all the observed kinematical features as well as the signatures of strong shock, whereas it assumed the presence of invisible objects (DMSH and BM).  

\begin{deluxetable}{lcc}[htbp]\label{table:smith}
\tablecaption{Comparison between the BM and Smith cloud }
\tablewidth{0pt}
\tablehead{
\colhead{Parameter} & \colhead{BM} & \colhead{Smith cloud} 
}
\startdata
Size/pc  & 15 & $3000\! \times\! 1000$ \\
Gas mass/$M_{\sun}$  & $\lesssim\! 6\!\times\! 10^{3}$ & $>\! 4\!\times\! 10^6$ \\
Velocity/\kms\ & $\sim\! 250$ & $\sim\! 300$ \\
Kinetic energy/$\mbox{erg}$ & $\lesssim\! 7\!\times\! 10^{51}$ & $>\! 4\!\times\! 10^{54}$ \\
DMSH size/pc  & $15\mbox{--}70$ & $>\! 3000$ \\
DMSH mass/$M_{\sun}$  & $\sim\! 6\!\times\! 10^7$ & $>\! 10^8$ \\
\enddata
\end{deluxetable}

According to this scenario, the clump of baryonic matter (BM), which is responsible for the formation of the CO shell, should measure $\sim\! 15$ pc, and predominantly consist of atomic hydrogen with an internal pressure comparable to that of molecular clouds, $p/k\!\gtrsim\! 10^3$ cm$^{-3}$ K.  Limited angular resolution of \Hone\ surveys ($16\farcm 2$ for HI4PI) and widespread \Hone\ emission in the Galactic plane prevents the direct detection of the BM in the CO shell.  We assume that the velocity width of the CO shell ($V_{\rm CO}\!\simeq\! 30$ \kms ) roughly represents the line-of-sight (LOS) component of the plunging velocity, and that its longitudinal component is similar to the LOS component ($|V_{l}|\!\simeq\! |V_{\rm LOS}|$).  Using the longitudinal and latitudinal extents of the \Hone\ filament ($\Delta l\! :\! \Delta b\! =\! 1\! :\! 4$), we roughly estimated the latitudinal component of the plunging velocity as $|V_{b}|\!\simeq\! 4\!\times\!|V_{l}|\! =\! 120$ \kms .  Thus the total plunging velocity of the BM/DMSH relative to the Galactic disk can be estimated as $V_{\rm BM} \! =\! \sqrt{V_{l}^2\! +\! V_{b}^2\! +\! V_{\rm LOS}^2} \!\simeq\! 130$ \kms .  Considering the Galactic rotational velocity at the CO shell ($\simeq 220$ \kms), the total plunging velocity with respect to the Galactic center is $\simeq\! 250$ \kms , which is a reasonable value for the Galactic halo objects \citep[e.g., ][]{Lockman02}.  Since the BM is left far behind the DMSH, it may has almost been stopped by the CO shell.  The conservation of momentum of the BM+CO shell system, $M_{\rm BM} V_{\rm BM}\!\lesssim\! (M_{\rm BM}\! +\! M_{\rm CO}) V_{\rm CO} $, provides a mass estimate of the BM approximately $M_{\rm BM}\!\lesssim\! 6\!\times\! 10^{3}\,M_{\sun}$.  The estimated parameters of the BM and Smith cloud are summarized in Table \ref{table:smith}, which demonstrates a sharp discrepancy in the scale between these two objects.

The size of the DMSH should fall in the range from $15$ pc to $70$ pc, corresponding to the diameter of the CO shell and the thickness of the \Hone\ filament (see Figure \ref{fig6}).  Equating the size of the \Hone\ void with twice of the Hoyle-Litteleton radius [eq.(6) in \citet{Edgar04} with $v_{\infty}\!=\! 130$ \kms ], we can roughly estimate the mass of the DMSH as $\sim\! 6\!\times\! 10^7\, M_{\sun}$.  This value lies at the middle of the theoretically hypothesized ``small" DMSH mass range of $10^{6\mbox{--}9}\, M_{\sun}$ \citep[e.g., ][]{Bovy,Ga}.  The detection and quantification of such small DMSHs is one of the most important goals of the $\Lambda$CDM cosmology \citep[][]{Bullock17}.  It is expected that the high-velocity plunge of such an object, which makes a \Hone\ void and filament, would have left certain signatures in the Galactic disk stellar population as well.  The detection of a ``vertical-velocity-field anomaly" toward the root of the \Hone\ filament in the Gaia astrometric data will be presented in a separate paper (K. Udagawa et al. 2024, in preparation).

\section{Summary}
We have performed mapping observations toward the BVF CO 16.134--0.553 in the CO {\it J}=1--0 and SiO {\it J}=2--1 lines using the NRO 45 m telescope.  The results are summarized as follows.  
\begin{enumerate}
\setlength{\parskip}{0pt} 
\setlength{\parsep}{0pt} 
\setlength{\itemsep}{1pt} 
\item The new CO maps showed that the 5-pc size BVF bridges two separate velocity components at $\textit V_{\rm{LSR}}\! \simeq\! 40$ \kms\ and $65\ \rm{km\ s^{-1}}$ in the position-velocity space.  
\item Prominent SiO emission was detected from the BVF and its root in the $\textit V_{\rm{LSR}}\! \simeq\! 40$ \kms\ component, indicating the passage of strong interstellar shock.    
\item Referring to the large-scale CO data, CO 16.134--0.553 appeared to correspond to the Galactic eastern rim of a 15-pc diameter expanding CO shell.  
\item An 70 pc-diameter \Hone\ void and 280 pc-long vertical \Hone\ filament were also found above and below the CO shell, respectively. 
\item A DMSH plunging scenario that has formed the \Hone\ void, CO 16.134--0.553/CO shell, and \Hone\ filament was proposed. 
\end{enumerate}

\begin{acknowledgments}
This study is based on observations at the Nobeyama Radio Observatory (NRO). We are grateful to the NRO staff for facilitating the operation of the telescope.  The NRO is a branch of the National Astronomical Observatory of Japan, National Institutes of Natural Sciences. We used the FUGIN data\footnote{https://nro-fugin.github.io/release/}. The data were retrieved from the JVO portal\footnote{http://jvo.nao.ac.jp/portal/} operated by ADC/NAOJ.  T.O. acknowledges the support from JSPS Grant-in-Aid for Scientific Research (A) No. 20H00178.  
\end{acknowledgments}

\end{document}